# Forecasting seasonal rainfall in SE Australia using Empirical Orthogonal Functions and Neural Networks

Stjepan Marčelja

Research School of Physics, Australian National University
Canberra, ACT 2601, Australia

*Correspondence to*: Stjepan Marčelja (stjepan.marcelja@anu.edu.au)

**Abstract.** Quantitative forecasting of average rainfall into the next season remains highly challenging, but in some favourable isolated cases may be possible with a series of relatively simple steps. We chose to explore predictions of austral springtime rainfall in SE Australia regions based on the surrounding ocean surface temperatures during the winter. In the first stage, we search for correlations between the target rainfall and both the standard ocean climate indicators as well as the time series of surface temperature data expanded in terms of Empirical Orthogonal Functions (EOFs). In the case of the Indian Ocean, during the winter the dominant EOF shows stronger correlation with the future rainfall than the commonly used Indian Ocean Dipole. Information sources with the strongest correlation to the historical rainfall data are then used as inputs into deep learning artificial neural networks. The resulting hindcasts appear accurate for September and October and less reliable for November. We also attempt to forecast the rainfall in several regions for the coming austral spring.

## 1 Introduction

Southeast Australia, including the Murray Darling basin, is a highly productive agricultural region largely dependent on adequate rainfalls providing irrigation water needed for high-value crops. Accurate prediction of water availability several months ahead would greatly aid in crop planning and management.

The most important influence on the future rainfall in Australia are ocean temperatures surrounding the continent. For more than 50 years Indian Ocean sea surface temperatures (SST) have been known to affect the rainfall over the continent. A comprehensive history of the earlier research is available in Ummenhofer et al. (2008). The configuration of the Indian Ocean SST is accepted as one of the major climate drivers over the large regions of the three continents at its boundaries. These SST conditions are codified as the Indian Ocean Dipole (IOD) index, defined as the difference between the anomaly at the western side (50ºE to 70ºE and 10ºN to 10ºS) and the eastern side (90ºE to 110ºE and 0ºS to 10ºS) of the basin. When in its negative phase, westerly winds at tropical latitudes accumulate warmer surface water north of Australia leading to increased rainfall over the continent.

The influence of Indian Ocean teleconnections on the rainfall in SE Australia were explored by Cai and Cowan (2008), Ummenhofer et al. (2009) and Cai et al. (2011) who established another indicator of the SST configuration of the Indian Ocean to the north-west of Australia that has an even stronger influence than the IOD on the rainfall in the south-east of the continent. The *Meridional temperature gradient*, defined in Ummenhofer et al. (2009) as the difference of SST anomalies between the regions *sI* (centered at 95ºE, 30ºS) and *eI* (centered at 110ºE, 10ºS), both with a 10º surrounds, strongly correlates with the rainfall during both dry and wet years. Closely linked to Indian Ocean conditions, subtropical ridge position and intensity are also strongly correlated with the rainfall in SE Australia (Cai *et al*. 2011a, 2011b; Timbal and Drosdowsky 2012).

Pacific Ocean conditions, as reflected in the phase of the El Niño Southern Oscillation (ENSO), influence the weather throughout Australia. The relationship of ENSO with weather systems movements and ultimately the rainfall in SE Australia is very variable with complicated dynamics (Lim et al. 2016a; Hauser et al. 2020). While ENSO and IOD are not independent (Wang et al., 2019), ENSO data contain information complementary to that of IOD and are the second most important input in forecasting rainfall in Australia.

Over decades, Pacific and Indian ocean climate patterns slowly change (Han et al., 2014; McKay et al., 2023a). Collected data become less relevant with the passage of time. In this work we found that pre-1950 data seldom contribute useful information, and for all hindcasting tests best results were obtained with the data sets starting at the year 2000.

The variations in the ocean conditions surrounding Australia are most prominent during the austral winter and spring, when both IOD and ENSO show larger departures from the neutral state (e. g. Lim et al., 2022). During the winter, this variability creates favourable conditions for medium range forecasting of the spring rainfall. While many of the physical processes driving climate pattern variability are understood we do not involve the physical basis of specific climate patterns. The physical processes are only considered as the sources of data that might be useful for the practical task of medium-term rainfall forecasting.

In the following section we look closely at the correlation between different climate drivers and the rainfall in the spring season. The three regions explored are, as defined by the Australian Bureau of Meteorology: Southeastern Australia, Victoria and the Murray Darling Basin (http://www.bom.gov.au/cgi-bin/climate/change/timeseries.cgi). In addition to other climate drivers, we introduce here the novelty of also using the expansion of ocean surface temperature distributions into EOFs. In the expansion, the full history of surface temperatures contained in the original data is represented as the mutually independent time series of expansion coefficients in decreasing order of importance. The results indicate that two added sources in forecasting, the Meridional gradient and dominant EOFs greatly improve the outcomes in all hindcast testing. Next section describes the search for best input data streams, and the following section describes the hind- and forecasting of the future rainfall with deep learning non-linear neural networks.

**2 Search for climate variables that correlate with future rainfall**

The primary search tool is the correlation (Pearson correlation coefficient), used to measure the linear relationship between the temperatures and rainfall at different locations. Most of the earlier work explored correlation between driving variables correlated with the rainfall and rainfall at the same time of the year. Here we are primarily interested in the predictive power of such variables, hence we search for the correlation between a variable and the rainfall in the following season. As an example, the correlation coefficient between the time series of August SST averages and the November rainfall in Southeastern Australia is shown in Fig. 1.

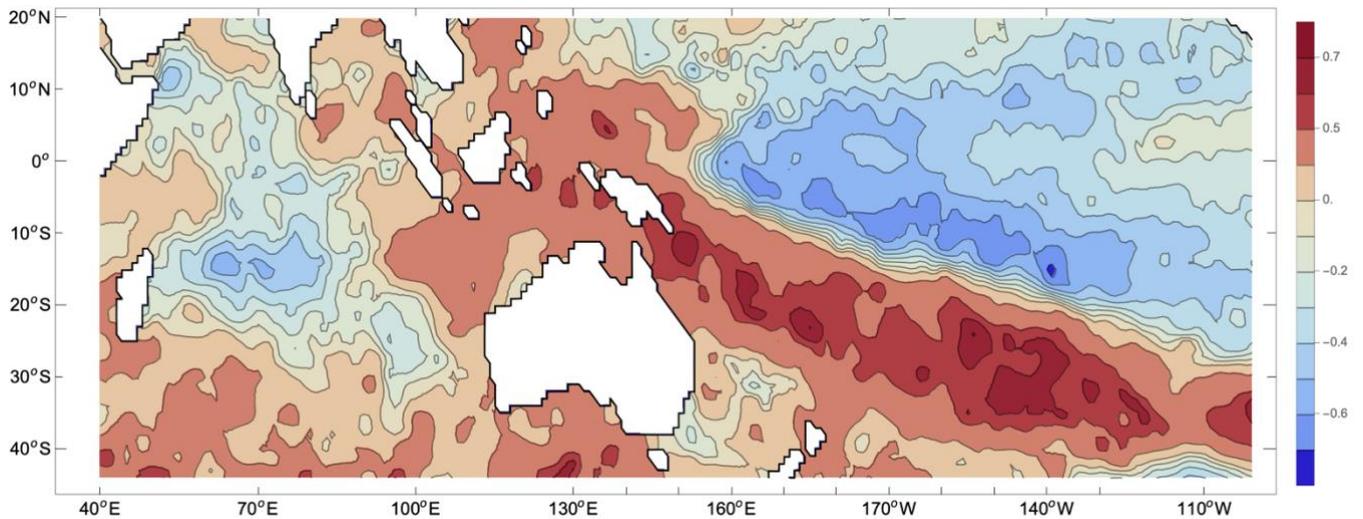

**Figure 1.** Correlation coefficient between the average SST during August and rainfall in November in Southeast Australia during the period 2000-2023, data set is HadISST4.01. The strong feature seen east of the Australian continent contains information useful for the forecasting of November rains.

A rigorous linear tool able to capture significant ocean SST features is the representation of SST fields as an expansion into the EOFs. The intent behind our use of EOFs as a forecasting tool is the reduction of the large and partly redundant data contained in SST at each position into smaller sets of independent data ordered in descending importance. EOFs were determined using the method and the software of Dawson (2016). For example, ocean surface temperatures at each position during the month of July between 1950 and 2024 are used as an input resulting in the time series of expansion coefficients for each EOF during the same period. The information in this condensed form leads to simpler and more accurate analysis.

The dominant July EOF for the Indian Ocean, shown in Fig. 2, is spatially similar to the definition of IOD. During July, the time series of its strength (expansion coefficient) correlates with rainfall in the September/October period stronger than the IOD. For example, the correlation coefficient between the Murray Darling basin rainfall during September/October period and the dominant Indian ocean EOF in July is 0.66. In this example the next best indicator is the Meridional gradient with the correlation coefficient of 0.55, while for IOD it is 0.44. The seemingly modest differences are very significant in hindcast testing of predictability.

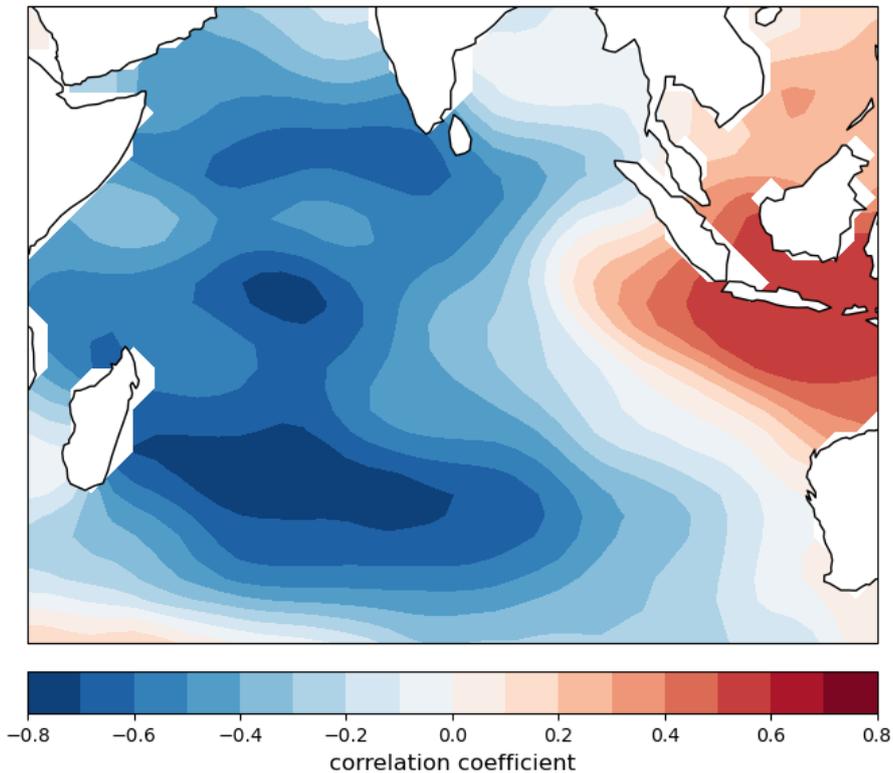

**Figure 2.** First EOF for the month of July and the Indian Ocean region shown in the figure expressed as a correlation. The expansion in EOFs is calculated from the time series of July SSTs between 1950 and 2024 using ERSSTv5 data.

After suitable ocean variables are identified, an easy way to check leading times between the SST signal and the subsequent rainfall is to plot correlation between monthly averages of the variable and the rainfall in each region. In Fig. 3 we illustrate the correlation coefficient between Indian Ocean variables and the rainfall in Southeast Australia over all months.

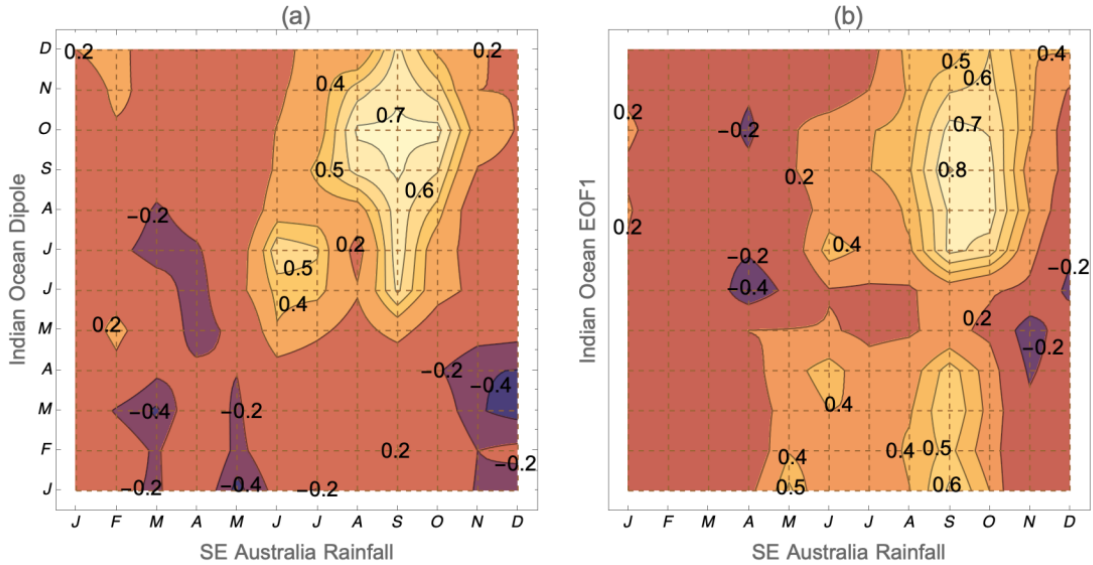

**Figure 3.** Correlation coefficient between the rainfall in Southeast Australia in any month and (a) IOD and (b) dominant Indian Ocean EOF during another month. The interesting part of the diagrams are the months preceding the rainfall, which are below the bottom left-top right diagonal. The images show that only in the springtime is rain strongly correlated with ocean indicators. It is easily visible that the Sept/Oct rain is more strongly correlated with the EOF than the IOD in July, while the opposite is true in June.

Finally, we evaluate *mutual information* (Cover and Thomas, 1991) which is a standard non-linear measure of information shared between two variables. In the present case, mutual information between an ocean variable at the time $t_1$, $O(t_1)$, and the rainfall at a different time $R(t_2)$ is defined as

$MI(O; R) = \Sigma_{t_1,t_2} \; P_{OR}(t_1, t_2) \log[P_{OR}(t_1, t_2)/(P_O(t_1) P_R(t_2))]$.

The discrete individual distributions $P_O(t_1)$ and $P_R(t_2)$ and the joint distribution $P_{OR}(t_1, t_2)$ were approximated by fitting the data to continuous Gaussian distributions and the double sum was then evaluated as a double integral over the fitted distributions. Typical results are presented in Fig 4, which shows the mutual information between the Meridional gradient and Southeast Australia rainfall at all times of a year.

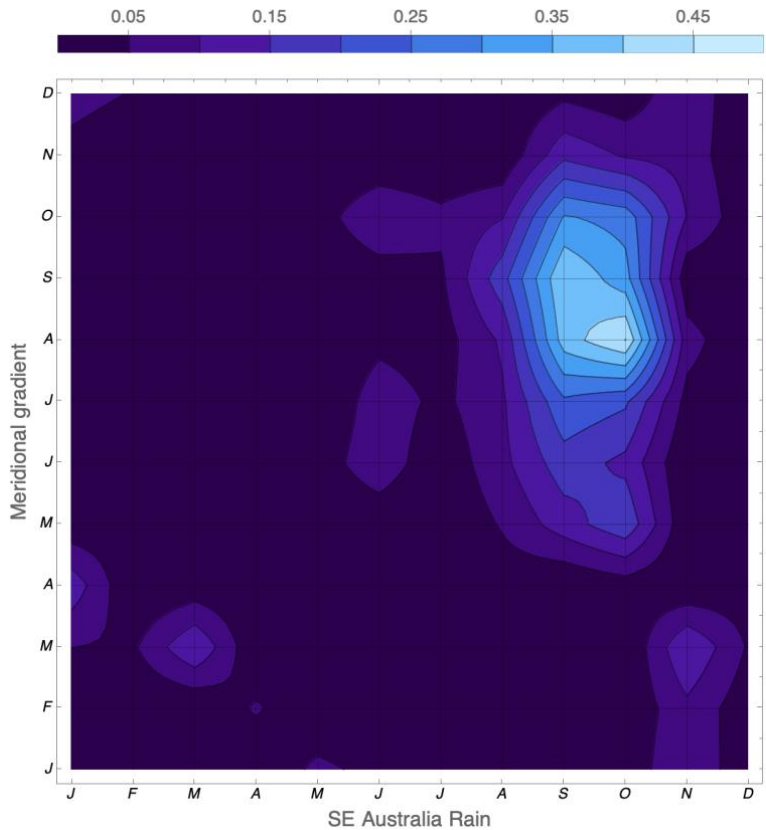

**Figure 4.** Mutual information as a non-linear example of a link between the ocean and rainfall variables. Only in spring is the joint information between the SE Australia rainfall and the earlier Meridional gradient significant, with the highest value of 0.45 obtained between Meridional gradient in August and SE Australia rainfall in October.

**3 Hind- and forecasting of springtime rainfall**

Once the information sources about future rainfall have been identified there is a choice between a number of possible linear and non-linear forecasting methods. The task is difficult, as regional climate changes over decades (Han et al., 2014) make older data misleading and the useful history period short. We found that the best results in hindcasting are obtained using only the records obtained since the year 2000. Data from 1950 were used in calculations of EOFs but not elsewhere.

Linear methods are less sensitive to errors in the input, and multiple linear regression is often a method of choice (e. g. Maher and Sherwood, 2014; Hudson et al., 2017; Lim et al., 2021; McKay et al., 2023b). In prediction of chaotic dynamics nonlinear methods, usually either delay coordinate embedding or neural networks, perform better (Weigend and Gershenfeld, 1993). In the period since 2000 there is not enough data to find similar past states and delay coordinate embedding is not accurate. In our tests with springtime rainfall deep learning artificial neural networks performed better than linear methods or delay coordinate embedding. Neural networks have been used in climate research on many occasions, e. g. El Niño or IOD forecasting (Nooteboom et al. 2018; Ham et al., 2019; Ratnam et al., 2020). More recently, very large-scale artificial intelligence models were developed for weather forecasting over a 10-day period (e. g. Lam et al., 2023; Bi et al., 2023) but they also have intrinsic limitations on accuracy (Seltz and Craig, 2023).

After extensive testing, we selected networks consisting of 3-5 linear layers and one tanh nonlinearity. 25 years of data were typically divided in about 14-17 years for training, 5-7 years for validation and one or more years for hindcasting. Networks were started from a random configuration. As is common with neural network training, the procedure is largely a matter of trial, error and gradual improvement (Hastie et al. 2009; Yong, 2022). Low learning rates lead to a slow drift towards one of the minima in the error landscape, while high learning rates allow the system to probe many of the neighbouring minima leading to a scatter in the results but often higher accuracy using the mean value. The input time series should be scaled to have the same, or almost the same norm, as relative magnitudes affect the non-linear process of training.

During neural networks training, a large number of synaptic weights are adjusted by fitting the available short sequence of target rainfall data. The networks rapidly drift towards the overfitting regime and the training is terminated when validation errors start to increase. To make the best use of the limited data, for the final forecast the validation segment is extended to the last available year and the usual test segment is omitted. Before this practice is adopted, it is necessary to check that the networks formed with only the data available a few years before the present still perform similarly to the networks trained using all the data. In cases when forecasting is robust, variations in the division between validation and test segments did not substantially change the results. The years when the networks fail in validation or hindcasting are always the same regardless of the details. An example of forecasting with and without a test segment is shown in Fig. 5. Terminating the validation segment in the year 2019 only slightly decreased the fit over the 2020-2023 period and practically left the forecast unchanged.

Within the validation and hindcasting test periods, from about 2015 onwards, the three regions explored (Victoria, Murray-Darling Basin and Southeast Australia) share all the main features of the rainfall. During September and October, rainfall trends and anomalies appear to follow a single trend, possibly pointing to a common physical origin. Using June, or better July data, hindcasting is fairly accurate, and we combine the two months into a single spring period. The behaviour and likely physical basis for November forecasting is different, and the predictions are less accurate. All rainfall anomalies were defined with respect to the 1991-2020 period, with the combined September/October rainfall averages of 114.8 mm, 73.4 mm and 108.5 mm for Victoria, Murray-Darling basin and SE Australia respectively. The corresponding November averages are 54.0 mm, 47.1 mm and 51.5 mm.

With June data, forecasting is based on both the well-known and not commonly used drivers, most important of the latter being the Meridional gradient. Other prominent inputs are Niño3.4 area SST and IOD. SST over other regions of the Indian Ocean is sometimes helpful, as are land temperatures over Australia in the case of the Murray-Darling region. The years of extreme rainfall are underestimated. The EOF expansion for June is useful but not an important source of information. Partial agreement in the validation period requires four to six input streams and the hindcasting success depends on details. While the results are sometimes unreliable, they still give a reasonable indication of the developing trends.

With July data, the best predictor is the first Indian Ocean EOF (region $40^0$E to $120^0$E and $40^0$S to $20^0$N). Adding just Niño 3.4 SST to the neural net inputs is already sufficient to provide good hindcasts. Further improvement is obtained by adding a few more input streams. The results using July data are robust and not sensitive to changes in the network structure or training details.

As a first example in Fig. 5 we show rainfall hindcasts and a forecast for the state of Victoria. When hindcasting with June data the extreme rainfall years are not fully anticipated. The learning rate during the training is set relatively high, resulting in a large scatter between individual runs, but also more accurate mean values. In Fig. 5b we show a test where the network is finalised with the data ending in 2017, thus reducing the validation set from 12 years to 8 years. The accuracy of the result is moderately reduced, but the main features are unchanged.

With July data it becomes possible to hindcast more extreme years but some of less important details are lost. Some years (2013 or 2020) are impossible to hindcast accurately, but the errors are not large.

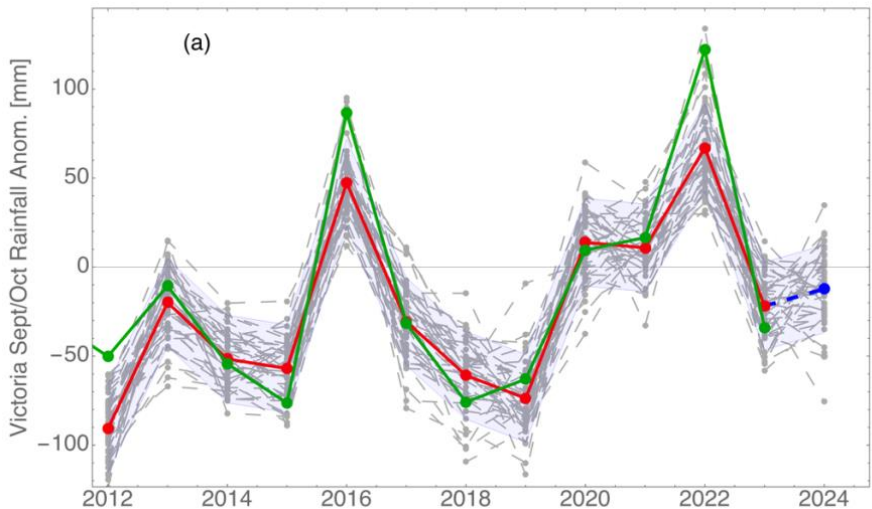

**Figure 5.**
Hind- and forecast of the rainfall anomaly in the state of Victoria during September and October using (a) June data; (b) June data with a test period (2020-2023) and (c) July data.

Historic rainfall is shown in green and mean values of the validation in red. The 2024 forecast is the last point in blue. Faint dashed lines are results from individual random network starts. Mean value of the RMS error over the period 2012-2023 is shown as shading. Training rainfall years were 2000-2011.

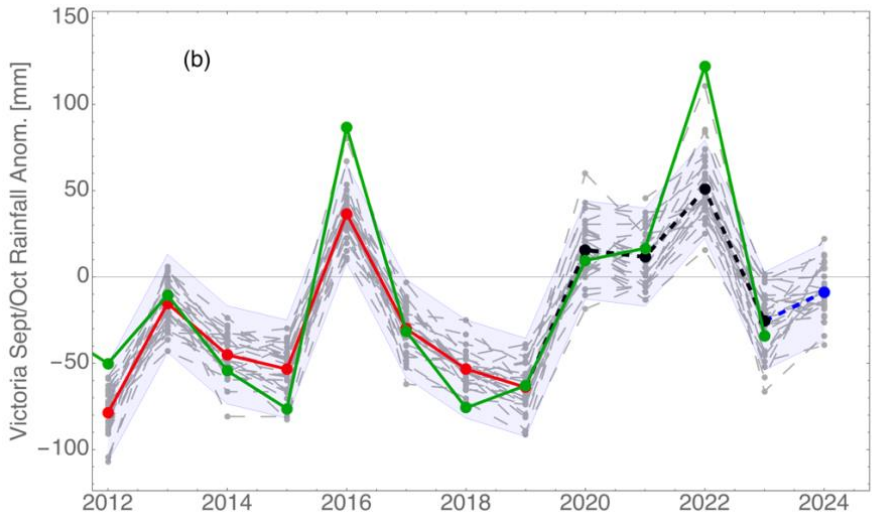

In (b), only the data up and including the year 2019 are used to define the network. Hindcast results using this network for years 2020-2023 are shown in black and the forecast for 2024 in blue.

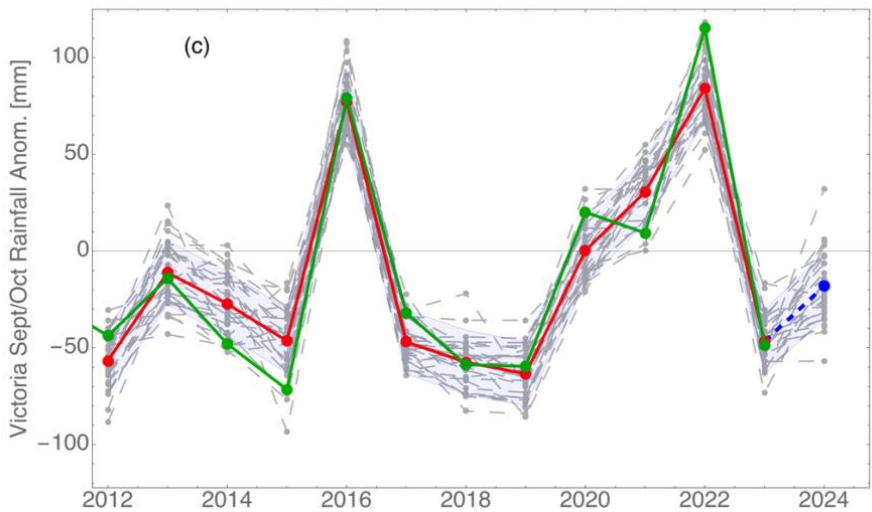

Turning next to the important Murray Darling basin the average errors are larger (Fig. 6) and the forecast is less reliable. Using the July data, the very large anomalies in 2016 and 2022 are hindcast correctly, but the problem years of 2013 and 2020 are still inaccurate. In addition to ocean data the input in (a) included temperatures on the Australian continent, while (b) is based on the first Indian Ocean EOF and Niño3.4 data only.

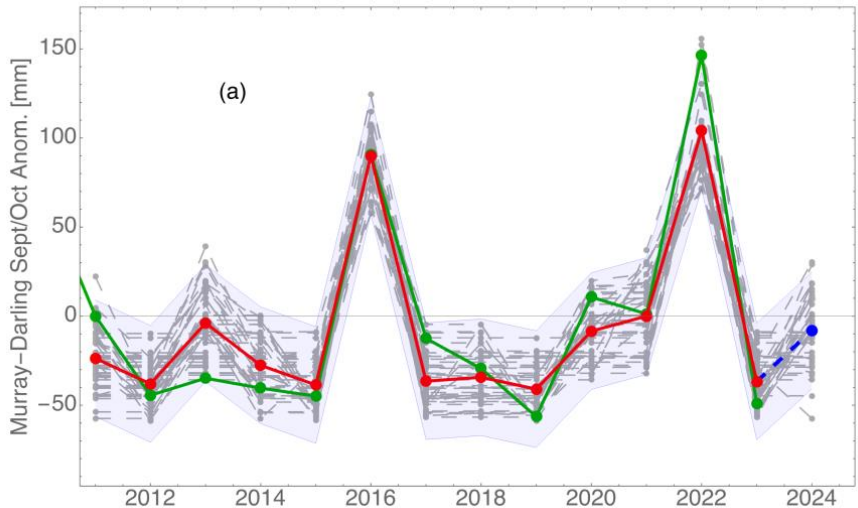

**Figure 6.**

Murray Darling basin Sept./Oct. rainfall hind- and forecast using (a) June data or (b) July data. The symbols have the same meaning as in Fig. 5.

In (a), setting a larger value for the network learning rate results in the system probing many states in search for the best solution. In (b), validation results were better with a low learning rate, when the system drifts to a local minimum without probing many adjacent states.

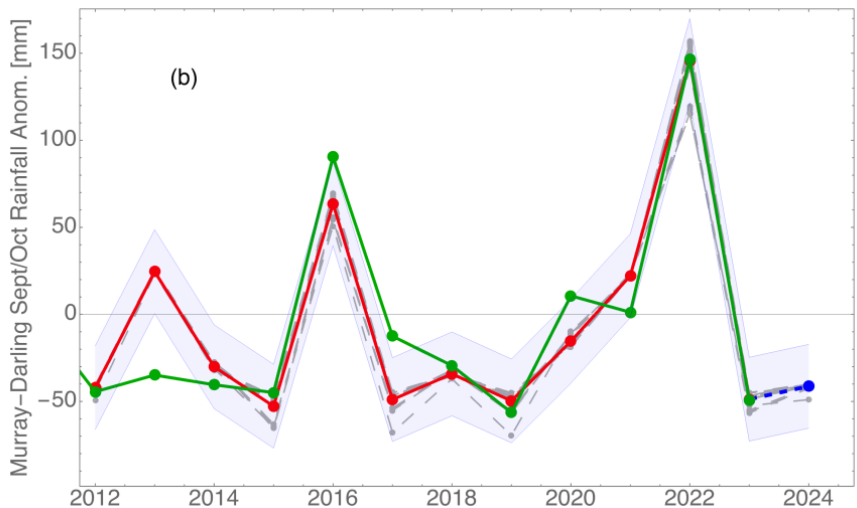

Hindcasting is most accurate for the larger SE Australia region (Fig. 7). June forecast shown in (a) is based on Niño3.4 SST, Meridional gradient, Australian continent temperatures and Murray-Darling temperatures. July forecast (b) is based on Indian ocean EOF1 and Niño3.4 SST. The hindcasts are similar in skill but the errors appear at different places. In (a), peaks are underestimated, while in (b) years 2013 and 2020 are incorrect.

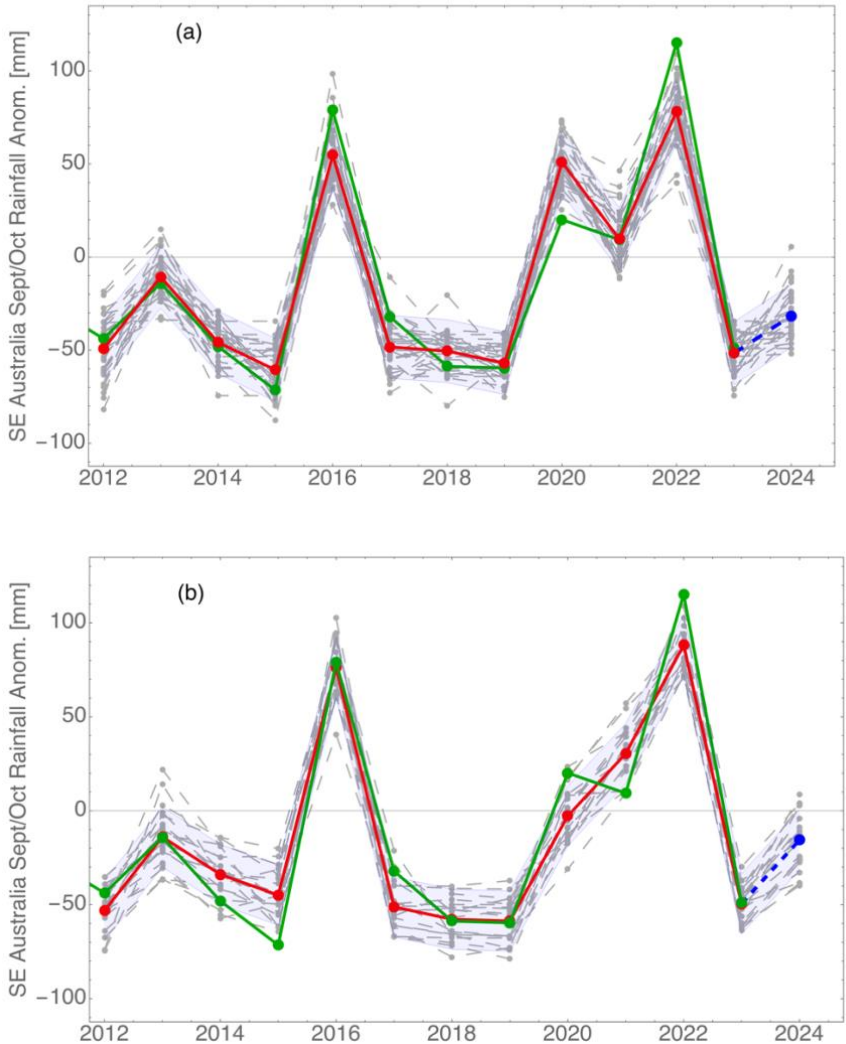

**Figure 7.** SE Australia September/October rainfall hind- and forecast using (a) June data or (b) July data. The symbols have the same meaning as in Fig. 5.

Lastly, we consider the rainfall in November hindcast with the August data. For input data, the strong south Pacific Ocean feature visible in Fig. 1 is included as the second EOF (region $140°E$ to $100°W$ and $40°S$ to $5°N$) together with several positions along the anomaly. In addition, the input includes Niño3.4 SST, the Indian Ocean Meridional gradient and the air surface pressure anomaly in Tahiti. The result shown in Fig 8 appears reasonable but only future forecasting will show if this combination is accurate and stable.

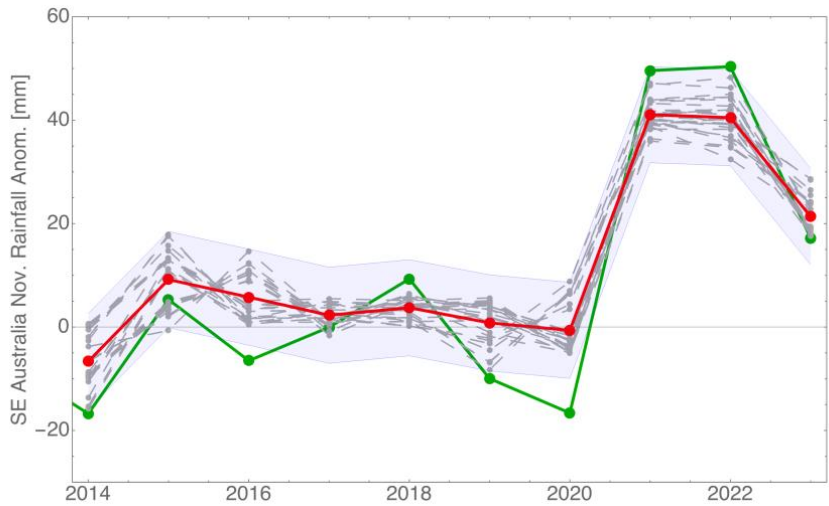

**Figure 8.**
Hindcast of 2023 November rainfall in SE Australia evaluated with August data. The symbols have the same meaning as in Figure 5.

**4 Discussion**

Ideally, throughout a year, a forecasting method should provide reliable forward information with only small errors, since perfect agreement is not possible due to the stochastic nature of weather events. The search described in Sec. 2 found strong correlations of different early indicators with the rainfall during the spring months and weaker correlation at other times of the year. For rainfall during winter months, correlations with SST, with accepted climate drivers, or with temperatures over different regions of the continent are weaker, with the correlation coefficient in the range 0.2-0.4. This information is insufficient with the network architectures tested here, and forecasts had to be restricted to spring months. Data with a high correlation coefficient of about 0.70-0.75 (when present with the EOF dominant principal component) are much more valuable than the data with the correlation coefficient of 0.60-0.65 (typical ocean indicators). To extend the forecasting to other periods of the year, advances are needed in using data with lower correlation to rainfall, perhaps with the increased range of inputs.

Within the accessible months, during the training of neural networks we encountered the general problem of overfitting the model when data sets are small, and many adjustable parameters are available. This problem has a human analogue, which may arise when the investigator has a choice between many available possibilities for the input data and input parameters, eventually finding the set that fits in hindcasting tests. A situation where only a few inputs are sufficient are preferable to hindcasting fits with up to ten different inputs. Since the selection of input data and training parameters has a large effect on the outcome, many tests were performed to explore the possible influence of overfitting resulting in unrealistic confidence.

Our concern was largely allayed when forecasting with very different sets of inputs in the June/July data comparisons still lead to similar outcomes.

Nonlinear neural networks used in this work performed well only in a very limited time and geographical window. At such times, hindcast testing suggests forecasts may be quantitatively accurate in difficult years of large anomalies, like 2016 or 2022, which would be an advantage over linear methods. At times other than the springtime months, when significant levels of correlation are still present, linear methods, like multi-linear regression are better (McKay et al., 2023b).

The neural networks application used here is an example of unsupervised learning, where no information is provided on the physical significance or the importance of each input data stream. In the longer term, we need to understand the physical mechanism driving climate fluctuations on the annual and seasonal scales. It is hoped that the present identification of features that have a significant influence on the rainfall during the following seasons may improve understanding of the physical basis of the relationships between the important climate variables and thus advance medium term climate forecasting.

COMPETING INTERESTS

The contact author declares no competing interests.

DATA SOURCES

*Rainfall and temperatures:* Australian Bureau of Meteorology
http://www.bom.gov.au/cgi-bin/climate/change/timeseries.cgi
*Niño34*: NOAA
https://psl.noaa.gov/data/climateindices/list/
*Indian Ocean Dipole (DMI):* NOAA
https://psl.noaa.gov/gcos_wgsp/Timeseries/Data/dmi.had.long.data
*Meridional Gradient, EOFs, selected ocean surface temperature regions were evaluated from Had or ERSST v5 ocean surface data using PyFerret.*
ERSST v5: http://iridl.ldeo.columbia.edu/SOURCES/.NOAA/.NCDC/.ERSST/.version5/.anom/datafiles.html
HadISST (delayed by several months)  https://www.metoffice.gov.uk/hadobs/hadisst/data/download.html

SOFTWARE AND MODEL CODE

*Ocean SST selections, nc files handling:* NOAA PyFerret
https://ferret.pmel.noaa.gov/Ferret/downloads/pyferret
*EOFS Evaluations:* Andrew Dawson EOF Analysis in Python
https://ajdawson.github.io/eofs/latest/
*Correlation maps and neural network evaluations: Mathematica* 14.0
https://www.wolfram.com/mathematica/
Author's brief *Mathematica* notebooks are available upon request.

**References**

Abram, N. J., Gagan, M. K., Cole, J. E., Hantoro, W. S. and Mudelsee, M.: Recent intensification of tropical climate variability in the Indian Ocean. Nature Geoscience 1, 849-853, doi:10.1038/ngeo357, 2008.

Bi, K., Xie, L., Zhang, H., Chen, X., Gu, X. and Tian, Q.: Accurate medium-range global weather forecasting with 3d neural networks. Nature, 619(7970), 1–6, doi:10.1038/s41586-023-06185-3, 2023.

Cai, W. and Cowan, T. Dynamics of late autumn rainfall reduction over southeastern Australia. Geophys. Res. Lett. L09708, doi:10.1029/2008GL033727, 2008.

Cai, W., van Rensch, P., Cowan, T. and Hendon, H. H.: Teleconnection pathways of ENSO and the IOD and the mechanisms for impacts on Australian rainfall, J. Clim., 24, 3910–3923, doi:10.1175/2011JCLI4129.1, 2011a.

Cai, W., van Rensch, P., Cowan, T.: Influence of Global-Scale Variability on the Subtropical Ridge over Southeast Australia, J. Clim., 24, 6035–6053, doi:10.1175/2011JCLI4149.1, 2011b.

Cover, T. M. and Thomas, J. A.: Elements of Information Theory, John Wiley & Sons, Inc. doi:10.1002/047174882X, 2006.

Dawson, A.: eofs: A Library for EOF Analysis of Meteorological, Oceanographic, and Climate Data, J. Open Research Software, 4, e14, doi:10.5334/jors.122, 2016.

Han, W., Vialard, J., J. McPhaden, M. J., Lee, T., Masumoto, Y., Feng, M. and de Ruijter, W. P. M., Indian ocean decadal variability: A Review. Bull. Am. Meteor. Soc. 95, 1680-1703, doi:10.1175/BAMS-D-13- 00028.1, 2014.

Yoo-Geun Ham, Y. G., Kim, J. H. and Luo, J. J.: Deep learning for multi-year ENSO forecasts. Nature 573, 568-572, doi:10.1038/s41586-019-1559-7, 2019.

Hastie, T., Tibshirani, R. and Friedman, J.: The Elements of Statistical Learning. Springer New York, NY, doi:10.1007/978-0-387-84858-7, 2009.

Hauser, S., Grams, C.M., Reeder, M.J., McGregor, S., Fink, A.H., Quinting, J.F., A weather system perspective on winter–spring rainfall variability in southeastern Australia during El Niño. QJR Meteorol. Soc. 146, 2614–2633, doi:10.1002/qj.3808, 2020.

Hudson, D., Alves, O., Hendon, H., Lim, E.-P., Liu, G., Luo, J.-J., MacLaughlan, C., Marshall, A. G., Shi, L., Wang, G., Wedd, R., Young, G., Zhao, M., and Zhou, X.: ACCESS-S1: The new Bureau of Meteorology multi-week to seasonal prediction system. J. South. Hemisph. Earth. Sys. Sci. 67, 132–159, doi:10.22499/3.6703.001, 2017.

Khastagir, A., Hossain, I and Anwar, A. H. M. F.: Efficacy of linear multiple regression and artificial neural network for long-term rainfall forecasting in Western Australia, Meteorol. Atmos. Phys., 134, 69, doi:10.1007/s00703-022-00907-4, 2022.

Lam, R. et al.: Learning skilful medium-range global weather forecasting, Science 382, 1416–142, doi:10.1126/science.adi2336, 2023.

Lim, E. P, Hendon, H.H., Zhao, M. and Yin, Y.: Inter-decadal variations in the linkages between ENSO, the IOD and south-eastern Australian springtime rainfall in the past 30 years. Clim. Dyn., 49, 97–112, doi:10.1007/s00382-016-3328-8, 2016a.

Lim, E. P., Hendon, H.H., Hudson, D., Zhao, M., Shi, L., Alves, O. and Young G.: Evaluation of the ACCESS-S1 hindcasts for prediction of Victorian seasonal rainfall. Bureau Research Report No. 19, Bureau of Meteorology Australia (available from: http://www.bom.gov.au/research/research-reports.shtml), 2016b.

Lim, E. P., Hudson, D., Wheeler M. C., Marshall A, G., King, A., Zhu, H., Hendon, H. H., de Burgh-Day, C., Trewin, B., Griffiths, M., Ramchurn, A. and Young, G.: Why Australia was not wet during spring 2020 despite La Niña. Scientific Reports, 11, 18423, doi:10.1038/s41598-021-97690-w, 2021.

Maher, P. and Sherwood, S. C.: Disentangling the Multiple Sources of Large-Scale Variability in Australian Wintertime Precipitation. J. Clim. 27, 6377-6392, doi:10.1175/JCLI-D-13-00659.1, 2014.

McKay, R. C., Pepler, A., Gillett, Z. E., Boschat, G, Rudeva, I., Purich, A., Dowdy, A., Hope, P. and Rauniyar, S.: Can southern Australian rainfall decline be explained? A review of possible drivers. WIREs Clim Change 14:e820, doi.org/10.1002/wcc.820, 2023a.

McKay, R., Lynette Bettio, L, Wang, W., Ramchurn, A. and Hope, P.: Multi-Linear Regression to Explain Australian Climate Events, http://www.bom.gov.au/research/workshop/2023/posters/RoseannaMcKay_poster.pdf , 2023b.

Nooteboom, P. D., Feng, Q. Y. López, C., Hernández-García, E, and Dijkstra, H. A.: Using network theory and machine learning to predict El Niño. Earth Syst. Dynam., 9, 969–983, doi:10.5194/esd-9-969, 2018.

Ratnam, J. V., Dijkstra, H. A. and Behera, S. K.: A machine learning based prediction system for the Indian Ocean Dipole. Sci. Reports 10, 284, doi:10.1038/s41598-019-57162-8, 2020.

Selz, T. and Craig, G. C.: Can Artificial Intelligence-Based Weather Prediction Models Simulate the Butterfly Effect? Geophys. Res. Lett. 50, e2023GL105747, doi:10.1029/2023GL105747 (2023).

Timbal, B. and Drosdowsky, W.: The relationship between the decline of Southeastern Australian rainfall and the strengthening of the subtropical ridge. Int. J. Climatol. 33, 1021-1034, doi: 10.1002/joc.3492, 2012.

Ummenhofer, C. C., Sen Gupta, A., Pook, M. J. and England, M. H.: Anomalous Rainfall over Southwest Western Australia Forced by Indian Ocean Sea Surface Temperatures. J. Clim. 21, 5113-5134, doi:10.1175/2008JCLI2227.1, 2008.

Ummenhofer, C. C., Sen Gupta, A., Taschetto, A. S. and England, M. H.: Modulation of Australian Precipitation by Meridional Gradients in East Indian Ocean Sea Surface Temperature. J. Clim. 22, 5597- 5610, doi:10.1175/2009JCLI3021, 2009.

Wang, H., Kumar, A., Murtugudde, R., Narapusetty, B. and Seip, K. L.: Covariations between the Indian Ocean dipole and ENSO: a modeling study, Clim. Dyn. 53, 5743-5761, doi:10.1007/s00382-019-04895-x, 2019.

Weigend, A. S. and Gershenfeld, N. A., eds.: Time Series Prediction, Addison Wesley, Reading, MA, doi:10.4324/9780429492648, 1994.

Yong, C. Y.: Geometry of Deep Learning, Springer Verlag, Singapore, doi:10.1007/978-981-16-6046-7, 2022.

761, (2019).